\documentclass[aps,showpacs,amsmath,preprint,amssymb,12pt]{revtex4}

\usepackage{graphicx}

\begin{document}
\newcommand{\tatevec}[2]{\left(   \begin{array}{c}
                         #1 \\ #2 \end{array}\right) }
\newcommand{\yokovec}[2]{\left(   \begin{array}{cc}
                         #1 & #2 \end{array}\right) }
\newcommand{\mat}[4]{\left(   \begin{array}{cc}
                         #1&#2 \\ #3&#4 \end{array} \right) }
\newcommand{\matsan}[9]{\left(   \begin{array}{ccc}
                         #1&#2&#3 \\ #4&#5&#6  \\ #7&#8&#9 
                         \end{array} \right) }
\newcommand{\detmat}[4]{\left| \begin{array}{cc}
                         #1&#2 \\ #3&#4 \end{array} \right| }
\newcommand{\del}[1]{ \partial_{#1} }
\newcommand{\Gam}[3]{ \Gamma^{#1}_{\, \, #2 #3} }
\newcommand{\delp}[1]{ \partial'_{#1} }
\newcommand{\Gamp}[3]{ \Gamma'^{\,#1}_{\,\,\,\, #2\, #3} }
\newcommand{\bv}[1]{{\mbox{\boldmath $#1$}}}
\newcommand{\tbv}[1]{{\mbox{\boldmath $\tilde{#1}$}}}
\def\Lam{{\Lambda}}
\def\eq{{\,\,=\,\, }}
\def\al{{\alpha}}
\def\lam{{\lambda}}
\def\vphi{{\varphi}}
\def\eps{{\epsilon}}
\def\bZ{{\bar{Z}}}
\def\bw{{\bar{w}}}
\def\tilU{{\tilde{U}}}
\def\mE{{{\mathcal E}}}

\title{Relationship Between Solitonic Solutions of Five-Dimensional Einstein Equations}

\author{Shinya Tomizawa${}^{1}$, Hideo Iguchi${}^{2}$ and Takashi Mishima${}^{2}$}

\affiliation{${}^{1}$ Department of Mathematics and Physics, Graduate School of Science, Osaka City University, 3-3-138 Sugimoto, Sumiyoshi,
Osaka~558-8585,~Japan\\
${}^{2}$ Laboratory of Physics,~College of Science and Technology,~
Nihon University,\\ Narashinodai,~Funabashi,~Chiba 274-8501,~Japan
}

\date{\today}

\begin{abstract}
We give the relation between the solutions generated by the inverse scattering method and the B\"acklund transformation applied to the vacuum five-dimensional Einstein equations. In particular, we show that the two-solitonic solutions generated from an arbitrary diagonal seed by the B\"acklund transformation are contained within those generated from the same seed by the inverse scattering method.

\end{abstract}

\preprint{OCU-PHYS 252}
\preprint{AP-GR 36}
\maketitle


\section{Introduction}
Recently, the studies on higher dimensional black holes have much attention, since it has been predicted that they would be produced in a future linear collider~\cite{BHinCollider}. In particular, a stationary black hole solution of Einstein equation has an important role in that it is expected to describe the classical equilibrium state and Hawking emission from it is considered to give us the signal of black hole production in the linear collider. By a lot of studies on higher dimensional black holes, it has been clarified that they have much complicated structure than in four-dimensions~\cite{Topology}. 
 
  Some of higher dimensional stationary black hole solutions were found. As higher dimensional generalization of the Kerr black hole solution, the Myers-Perry black hole solution, which has an event horizon with spherical topology, was found~\cite{Myers:1986un}. Emparan and Reall found a black ring solution as the five-dimensional vacuum Einstein equation, which has an event horizon diffeomorphic to $S^1\times S^2$~\cite{Emparan:2001wn} and describes a black hole rotating in the $S^1$ direction. 

The black ring solution which is rotating only in $S^2$ direction was also found by two of the present authors~\cite{MI}
by using one of the solitonic solution-generating
techniques \cite{Castejon-Amenedo:1990b}, so-called B\"acklund transformation. 
This is essentially the technique to generate new solutions of the Ernst equation from a known solution. In addition, it was shown by same authors that the black ring with $S^1$ rotation
can be generated by the same solution generating technique \cite{MI2}.

As another solitonic technique, the inverse scattering technique~\cite{Belinskii} developed by Belinski and Zakharov is well known. This technique is essentially based on the fact that the Einstein's second-order nonlinear partial differential equations can be replaced  with
a pair of first-order linear partial differential
equations called Lax pair. This method produces vacuum solutions from a certain known vacuum solution called a seed and succeeded in generation of a lot of
four-dimensional solutions.
In fact, as four-dimensional solutions
the Kerr black hole solution, the multi Kerr black hole solutions
and the Tomimastu-Sato solutions can be generated from the Minkowski seed and a  physically interesting various solutions were also generated~\cite{textism,exact}.

Recently, some of higher-dimensional black hole/ring solutions
have been generated by using the inverse scattering method.
As an infinite number of static solutions of the five-dimensional
vacuum Einstein equations with axial symmetry,
the five-dimensional Schwarzschild solution
and the static black ring solution were reproduced \cite{Koikawa:2005ia},
which gave the first example of the generation of 
a higher-dimensional asymptotically flat black hole solution 
by the inverse scattering method.
The Myers-Perry solution with single and double angular momenta
were regenerated from the Minkowski \cite{Tomizawa,Koikawa:2005ia} and
some unphysical seed~\cite{Pomeransky:2005sj}, respectively.
The black ring solutions with $S^2$ rotation 
was also reproduced by using this method
from the Minkowski seed~\cite{Tomizawa}.
The $S^1$-rotating black ring  solution was also reproduced
from the Levi-Civita solution via the inverse scattering method by one of authors~\cite{Tomizawa2}.

Thus, the inverse scattering method is a powerful formalism for solving systems of nonlinear partial differential equations such as Einstein equation. When we try to find a further new higher dimensional black hole/ring solution, it is important to know the relationship between solutions generated by the B\"acklund transformation used in Ref.~\cite{MI} and ones generated by the inverse scattering method used in Ref.~\cite{Tomizawa,Tomizawa2} in that we can avoid the overlap of obtained solutions. Though the relationship between the four-dimensional solutions generated by the inverse scattering method and (the B\"acklund transformation was 
given) some B\"acklund transformations were considered in Ref.~\cite{G,G1}(see also Ref.~\cite{Cos,exact} about the relationship between the other 
generation-techniques), the relationship between higher dimensinoal solutions generated by these techniques dicussed here may be non-trivial and useful.
In usual cases, in these solitonic generation-techniques a diagonal metric are often used as a seed since such a seed simplify the analysis. These techniques have a merit in that more complex solutions can be generated from simple solutions. This is why in this article, we investigate the relation between the solutions generated from a diagonal seed by both solitonic generation-techniques applied to five dimensions, and show that the two-solitonic solutions generated by B\"acklund transformation in Ref.~\cite{MI} from an arbitrary diagonal seed coincide with two-solitonic solutions generated from the same seed by the inverse scattering method under the special normalization.

This article is organized as follows : In Sec.\ref{sec:pre}, we review the B\"acklund transformation developed by two of authors in Ref.~\cite{MI} and the inverse scattering method applied to five dimensions by one of authors~\cite{Tomizawa}. In Sec.\ref{sec:proof}, we show that the two-solitonic solutions generated by the former technique from a diagonal seed coincide with the special solutions generated by the latter technique from the same seed under the special normalization.


\section{Preliminary}\label{sec:pre}

In this article, we consider 
the spacetimes which satisfy the following conditions:
(1) five dimensions, (2) asymptotically flat spacetimes, 
(3) the solutions of 
vacuum Einstein equations, (4) having three commuting Killing vectors 
including one time-translational Killing vector  and two axial Killing vectors.
(5) having a single nonzero angular momentum component.


\subsection{B\"acklund transformation}\label{sec:back}
Here, we review the generation-technique of the five dimensional solution established by Ref.~\cite{MI}.
Under the conditions $(1)-(5)$, 
we can start the analysis from the following form of metric
\begin{eqnarray}
ds^2 &=&e^{-T}\left[
       -e^{S}(dt-\omega d\phi)^2
       +e^{T+2U_1}\rho^2(d\phi)^2 \right. \nonumber \\
&&\hskip 0cm \left.
+e^{2(\gamma+U_1)+T}\left(d\rho^2+dz^2\right) \right]
  +e^{2T}(d\psi)^2.
\label{eq:MBmetric}
\end{eqnarray}
Using this metric form
the Einstein equations are reduced to the following set of equations, 
\begin{eqnarray*}
&&{\bf\rm (i)}\quad
\nabla^2T\, =\, 0,   \\
&&{\bf\rm (ii)}\quad
{
\left\{\begin{array}{ll}
& \del{\rho}\gamma_T={
  \frac{3}{4}\,\rho\,
  \left[\,(\del{\rho}T)^2-(\del{z}T)^2\,\right]}\,\ \   \\[3mm]
& \del{z}\gamma_T={\displaystyle 
\frac{3}{2}\,\rho\,
  \left[\,\del{\rho}T\,\del{z}T\,\right],  }
 \end{array}\right. } \\
&&{\bf\rm (iii)}\quad
\nabla^2\mE_S=\frac{2}{\mE_S+{\bar\mE}_S}\,
                    \nabla\mE_S\cdot\nabla\mE_S , \\  
&&{\bf\rm (iv)}\quad
{
\left\{\begin{array}{ll}
& \del{\rho}\gamma_S={\displaystyle
\frac{\rho}{2(\mE_S+{\bar\mE}_S)}\,
  \left(\,\del{\rho}\mE_S\del{\rho}{\bar\mE}_S
  -\del{z}\mE_S\del{z}{\bar\mE}_S\,
\right)}     \\
& \del{z}\gamma_S={\displaystyle
\frac{\rho}{2(\mE_S+{\bar\mE}_S)}\,
  \left(\,\del{\rho}\mE_S\del{z}{\bar\mE}_S
  +\del{\rho}\mE_S\del{z}{\bar\mE}_S\,
  \right)},
\end{array}\right. } \\
&&{\bf\rm (v)}\quad
\left( \del{\rho}\Phi,\,\del{z}\Phi \right)
=\rho^{-1}e^{2S}\left( -\del{z}\omega,\,\del{\rho}\omega \right),  \\
&&{\bf\rm (vi)}\quad 
\gamma=\gamma_S+\gamma_T,   \\
&&{\bf\rm (vii)}\quad 
U_1=-\frac{S+T}{2},
\end{eqnarray*}
where the function $\Phi(\rho,z)$ is defined through the equation (v) and the function 
$\mathcal{E_S}$ is defined by 
$
\,\mE_S:=e^{S}+i\,\Phi\,.
$ It should be noted that $e^{S}$ and $\Phi$ corresponds to a gravitational potential and a twist potential. The equation (iii) is exactly the same as the Ernst equation in four dimensions \cite{refERNST}.
The most nontrivial task to obtain new metrics is to solve 
the equation (iii) because of its nonlinearity. 
Here the method similar to the Neugebauer's 
B\"{a}cklund transformation \cite{Neugebauer:1980} 
or the Hoenselaers-Kinnersley-Xanthopoulos transformation 
\cite{Hoenselaers:1979mk} is used. 

Following the procedure
given by Castejon-Amenedo and Manko \cite{ref7}, for a static seed solution $e^{S^{(0)}}$ a new Ernst potential can be written in the form
\begin{equation}
{\cal E}_S = e^{S^{(0)}}\frac{x(1+ab)+iy(b-a)-(1-ia)(1-ib)}
                         {x(1+ab)+iy(b-a)+(1-ia)(1-ib)}, \nonumber
\end{equation}
where $x$ and $y$ are the prolate-spheroidal coordinates:
$
\,\rho=\sigma\sqrt{x^2-1}\sqrt{1-y^2},\ z=\sigma xy\,
$
with the ranges $1\le x$ and $-1 \le y \le 1$,
and the functions $a$ and $b$ satisfy the following 
simple first-order differential equations 
\begin{eqnarray}
(\ln a)_{,x}&=&\frac{1}{x-y}
[(xy-1)S^{(0)}_{,x}+(1-y^2)S^{(0)}_{,y}], \nonumber \\
(\ln a)_{,y}&=&\frac{1}{x-y}
[-(x^2-1)S^{(0)}_{,x}+(xy-1)S^{(0)}_{,y}], \nonumber\\
(\ln b)_{,x}&=&-\frac{1}{x+y}
\left[(xy+1)S^{(0)}_{,x}+(1-y^2)S^{(0)}_{,y}\right], \nonumber\\
(\ln b)_{,y}&=&-\frac{1}{x+y}
\left[-(x^2-1)S^{(0)}_{,x}+(xy+1)S^{(0)}_{,y}\right]. \nonumber\\
&&    \label{eq:ab}
\end{eqnarray}
The corresponding expressions for the metric functions can be obtained
by using the formulas shown by \cite{ref7}.  
For the seed,

\begin{eqnarray}
ds^2 &=&e^{-T^{(0)}}\left[
       -e^{S^{(0)}}dt^2
       +e^{-S^{(0)}}\rho^2(d\phi)^2 \right. \nonumber \\
&&\hskip 0cm \left.
+e^{2\gamma_{(0)}-S^{(0)}}\left(d\rho^2+dz^2\right) \right]
  +e^{2T^{(0)}}(d\psi)^2.
\label{eq:backseed}
\end{eqnarray}
 a new solution is given by
\begin{eqnarray}
ds^2&=&-e^{S^{(0)}-T^{(0)}}\frac{A}{B}\biggl[dt-\biggl(2\sigma e^{-S^{(0)}}\frac{C}{A}+C_1\biggr)d\phi\biggr]^2\nonumber\\
    & &+\frac{B}{A}e^{-S^{(0)}-T^{(0)}}\sigma^2(x^2-1)(1-y^2)d\phi^2+e^{2T^{(0)}}d\psi^2\nonumber\\
    &&+ e^{2\gamma-S^{(0)}-T^{(0)}}\frac{B}{A}\sigma^2 (x^2-y^2)\biggl(\frac{dx^2}{x^2-1}+\frac{dy^2}{1-y^2}\biggr), \label{eq:sol}
\end{eqnarray} 
where $A$, $B$ and $C$ are defined by
\begin{eqnarray}
&&A:=(x^2-1)(1+ab)^2-(1-y^2)(b-a)^2,\nonumber\\
&&B:=[(x+1)+(x-1)ab]^2+[(1+y)a+(1-y)b]^2,\nonumber\\
&&C:=(x^2-1)(1+ab)[(1-y)b-(1+y)a]+(1-y^2)(b-a)[(x+1)-(x-1)ab].\label{eq:A-C}
\end{eqnarray}  
To assure that the spacetime does not have global rotation, 
the constant $C_1$ is give by
\begin{eqnarray}
C_1=\frac{2\sigma^{1/2}\alpha}{1+\alpha\beta}. 
\end{eqnarray}

\subsection{Inverse scattering Techniques}\label{sec:ism}

We will give a summary of the inverse scattering method
developed by Belinski and Zakharov \cite{Belinskii}
which is applied to five dimensions.

As in the previous subsection, we consider the asymptotically flat, five-dimensional
stationary and axisymmetric vacuum spacetime
with three commuting Killing vector fields $V_{(i)}$
($i=1,2,3$) following the argument in
\cite{Tomizawa,Tomizawa2}.
The commutativity of Killing vectors $[V_{(i)}, V_{(j)}]=0$
enables us to find a coordinate system such that
$V_{(i)}=\partial/\partial x^i~(i=1, 2, 3)$ and
the metric is independent of the coordinates $x^i$,
where $(\partial/\partial x^1)$ is the Killing vector field
associated with time translation and
$(\partial/\partial x^2),(\partial/\partial x^{3})$
denote the spacelike Killing vector fields with closed orbits.
We put $x^1=t,\ x^2=\phi,$ and $x^3=\psi$.
 From the theorem in Ref.~\cite{weyl},
in such a spacetime, the metric can be written
in the canonical form~\cite{weyl} as
\begin{eqnarray}
ds^2=f(d\rho^2+dz^2)+g_{ij}dx^idx^j,
\label{eq:canonical}
\end{eqnarray}
where $f=f(\rho,z)$ and $g_{ij}=g_{ij}(\rho,z)$
are a function and an induced metric on the
three-dimensional space, respectively.
Both of them depend only on the coordinates $\rho$ and $z$.
Here it is the most convenient to choose
the $3\times 3$ matrix $g=(g)_{ij}$ as to satisfy the condition
\begin{eqnarray}
{\rm det}g=-\rho^2.
\label{eq:det}
\end{eqnarray}
This is compatible with the vacuum Einstein equations
$g^{ij}R_{ij}=0$, which reduces to
$(\partial _\rho ^2+\partial _z^2)(-{\rm det}g)^{1/2}=0$.
It follows from $R_{ij}=0$ that the matrix
$g$ satisfies the solitonic equation
\begin{eqnarray}
(\rho g_{,\rho} g^{-1})_{,\rho}+(\rho g_{,z}g^{-1})_{,z}=0.
\label{eq:soliton}
\end{eqnarray}
>From the other components of the Einstein
equations $R_{\rho\rho}-R_{zz}=0$ and $R_{\rho z}=0$,
we obtain the equations which determine the function $f(\rho,z)$
for a given solution of the solitonic equation (\ref{eq:soliton})
\begin{eqnarray}
(\ln f)_{,\rho}&=&-\frac{1}{\rho}+\frac{1}{4\rho}{\rm Tr}(U^2-V^2),
\label{eq:f1} \\
(\ln f)_{,z}&=&\frac{1}{2\rho}{\rm Tr}(UV),\label{eq:f2}
\end{eqnarray}
where the $3\times 3$ matrices $U(\rho,z)$ and $V(\rho,z)$
are defined by
\begin{eqnarray}
U:=\rho g_{,\rho}g^{-1}, \qquad V:=\rho g_{,z}g^{-1}.
\end{eqnarray}
The integrability condition with respect to $f$ is automatically
satisfied for the solution $g$ of Eq.~(\ref{eq:soliton}).
Note also that $R_{\rho \rho }+R_{zz}=0$ is consistent with
the solution (\ref{eq:soliton}), (\ref{eq:f1}) and (\ref{eq:f2}).

Although our immediate goal is to solve the differential
equations (\ref{eq:soliton}), it cannot be generally solved
due to its non-linearity.
But in analogy with the soliton technique,
we can find the Lax pair for the matrix
equations (\ref{eq:soliton}).
We consider Schr\"odinger type equations
for the $3\times 3$ matrix $\psi (\lambda ,\rho ,z)$
as in four dimensions;

\begin{eqnarray}
D_{1}\psi =\frac{\rho V-\lambda U}
{\lambda ^2+\rho ^2}\psi,
\qquad
D_{2}\psi =\frac{\rho U+\lambda V}
{\lambda ^2+\rho ^2}\psi,
\label{eq:Laxpair}
\end{eqnarray}
where $\lambda $ is a complex spectral
parameter independent of $\rho $ and $z$.
The differential operators
$D_1$ and $D_{2}$ are defined as
\begin{eqnarray}
D_{1}:=\partial _z-\frac{2\lambda ^2}
{\lambda ^2+\rho ^2}\partial _\lambda ,
\qquad
D_{2}:=\partial _\rho +\frac{2\lambda \rho }
{\lambda ^2+\rho ^2}\partial _\lambda ,
\label{eq:LAeq}
\end{eqnarray}
which can be shown to commute $[D_1, D_2]=0$.
Note that Eq. (\ref{eq:LAeq}) is invariant under
the transformation $\lambda \to -\rho ^2/\lambda $.
Then the compatibility condition
$[D_1, D_2]\psi =0$ reduces to
the Einstein equations (\ref{eq:soliton}) with
\begin{eqnarray}
g(\rho ,z)=\psi (0, \rho, z).
\label{eq:initial}
\end{eqnarray}
It deserves to note that the Einstein's second-order
nonlinear partial differential
equations (\ref{eq:soliton}) are reduced to
a pair of first-order linear partial differential equations
(\ref{eq:Laxpair}).

Let $g_0, U_0, V_0$ and $\psi_0$ be
particular solutions of Eq. (\ref{eq:soliton})
and (\ref{eq:Laxpair}).
We shall call the known solution $g_0$
the seed solution.
We are going to seek a new solution
of the form
\begin{eqnarray}
\psi =\chi \psi_0,
\label{eq:DM}
\end{eqnarray}
which leads the following equations 
that the dressing matrix
$\chi (\lambda , \rho ,z)$ must satisfy 
\begin{eqnarray}
D_1 \chi =\frac{\rho V-\lambda U}
{\lambda ^2+\rho ^2}\chi
-\chi \frac{\rho V_0-\lambda U_0}
{\lambda ^2+\rho ^2},\nonumber \\
\label{eq:DMeq}\\
D_2 \chi =\frac{\rho U+\lambda V}
{\lambda ^2+\rho ^2}\chi
-\chi \frac{\rho U_0+\lambda V_0}
{\lambda ^2+\rho ^2}.\nonumber
\end{eqnarray}

In order for the solutions $g(\rho ,z)$
to be real and symmetric,
we impose the following conditions
on the dressing matrix $\chi$,
\begin{eqnarray}
\bar\chi (\bar \lambda ,\rho ,z)=\chi (\lambda ,\rho ,z),
\qquad
\bar\psi (\bar \lambda ,\rho ,z)=\psi (\lambda ,\rho ,z),
\label{eq:realg}
\end{eqnarray}
and
\begin{eqnarray}
g=\chi (-\rho ^2/\lambda ,\rho ,z)g_0
{}^T\hspace{-0.1cm}\chi (\lambda ,\rho ,z),
\label{eq:symg}
\end{eqnarray}
where $\bar \chi $ and ${}^T\chi$  denote
complex conjugation and the transposition of $\chi$.
>From Eqs. (\ref{eq:DM})
and (\ref{eq:symg}), the dressing matrix
$\chi$ asymptotes to a unit matrix
$\chi \to I$ as $ \lambda \to \infty $.

The general $n$-soliton solutions for the
matrix $g$ are generated due to the presence of the
simple poles of the dressing matrix on the complex
$\lambda $-plane:
\begin{eqnarray}
\chi =I+\sum_{k=1}^n\frac{R_k}
{\lambda -\mu _k},
\label{eq:DMform}
\end{eqnarray}
where the matrices $R_k$ and the position
of the pole $\mu _k$ depend only on
the variables $\rho $ and $z$.
Here and hereafter, the subscript $k, l$
counts the number of solitons.
It is the characteristic feature of solitons 
that the dressing matrix 
$\chi$ is represented as the meromorphic function
on the complex $\lambda$-plane.
Pole trajectories $\mu _k (\rho ,z)$
are determined by
the condition that the left-hand side of Eq.
(\ref{eq:DMeq}) have no poles of second-order
at $\lambda =\mu _k$, which yields following two
differential equations for $\mu _k (\rho ,z)$:
\begin{eqnarray}
\mu _{k,z}=-\frac{2\mu _k^2}
{\mu _k^2+\rho ^2},\label{eq:muz}
\qquad
\mu _{k,\rho }=\frac{2\rho \mu _k}
{\mu _k^2+\rho ^2},
\label{eq:eqmuk}
\end{eqnarray}
which are expressed by the solutions of the
following quadratic equations
\begin{eqnarray}
\mu _k^2+2(z-w_k)\mu _k-\rho ^2=0,
\label{eq:quadmu}
\end{eqnarray}
where $w_k$ are arbitrary constants.
Solving Eq. (\ref{eq:quadmu}), one can easily see
\begin{eqnarray}
\mu _k=w_k-z\pm \sqrt{(z-w_k)^2+\rho ^2},
\label{eq:mup}
\end{eqnarray}
where $w_k$ are arbitrary constants.
Since the matrices $R_k$ are degenerate
at the poles $R_k \chi^{-1}(\mu _k)=0$,
which follows from the condition $\chi \chi^{-1}=I$
at $\lambda =\mu _k$, it is possible to
write down the matrix elements of $R_k$
in the form
\begin{eqnarray}
(R_k)_{ij}=n_i^{(k)}m_j^{(k)}.
\end{eqnarray}
The fact that Eq. (\ref{eq:DMeq}) has no
residues at the poles $\lambda =\mu _k $ leads
to obtain the vectors $m_i^{(k)}$ as
\begin{eqnarray}
m^{i(k)}= m_{0j}^{(k)}[\psi_0^{-1}
(\mu_k, \rho ,z)]^{ji},
\label{eq:vectorm}
\end{eqnarray}
where $m_{0i}^{(k)}$ are arbitrary constants.
The vectors $n_i^{(k)}$, on the other hand,
are determined
by the condition that Eq. (\ref{eq:symg})
is regular at $\lambda =\mu _k$ as
\begin{eqnarray}
n_i^{(k)}=\sum _{l=1}^n\mu _k^{-1}(\Gamma ^{-1})
_{kl}L_{i}^{(l)},
\label{eq:vectorn}
\end{eqnarray}
where the vectors $L_i^{(k)}$ and
the symmetric matrix $\Gamma _{kl}$ are given by
\begin{eqnarray}
L_i^{(k)}&=&m^{j(k)}(g_0)_{ij},
\label{eq:vectorl}\\
\Gamma _{kl}&=&\frac{m^{i(k)}(g_0)_{ij}m^{j(l)}}
{\rho ^2+\mu _k\mu _l},
\label{eq:gamma}
\end{eqnarray}
respectively. Therefore one can now find
from Eq. (\ref{eq:initial}),
(\ref{eq:DM}) and (\ref{eq:DMform}) that
the matrix $g$ becomes
\begin{eqnarray}
g^{({\rm unphys})}_{ij}
&=&\psi (0, \rho ,z)_{ij}\nonumber \\
&=&(g_0)_{ij}-\sum_{k,l=1}^n(\Gamma^{-1})_{kl}\mu _k
^{-1}\mu _l^{-1}L_i^{(k)}L_j^{(l)}.
\label{eq:unphys1}
\end{eqnarray}
This metric does not meet the condition
${\rm det}g=-\rho^2$, which we have denoted
$g^{({\rm unphys})}$.
In order to satisfy the gauge condition
${\rm det}g=-\rho^2$,
the metric should be appropriately normalized.
One example is to normalize
all the metric components by the same weight as

\begin{eqnarray}
g^{{\rm (phys)}}=(-1)^{{n}/{3}}\rho^{-{2n}/{3}}
\left(\prod_{k=1}^n\mu_k^{{2}/{3}}\right)g^{{\rm (unphys)}},
\label{eq:nor}
\end{eqnarray}
where $g^{{\rm (phys)}}$ is the metric
which fulfills the condition ${\rm det}g=-\rho^2$.
Actually, the four-dimensional Kerr solution is obtained
similarly by the overall normalization as Eq. $(\ref{eq:nor})$.
Substituting the physical metric solution
$g^{{\rm (phys)}}$ given by Eq. (\ref{eq:nor})
into Eq. (\ref{eq:f1}) and  (\ref{eq:f2}),
we obtain a physical value of $f$ as

\begin{eqnarray}
f=C_0f_0\rho^{-{n(n-1)}/{3}}{\rm det}
(\Gamma_{kl})\prod_{k=1}^n\left[\mu_k^{{2(n+2)}/{3}}
(\mu_k^2+\rho^2)^{{-1}/{3}}\right]\cdot
\prod_{k>l}^n(\mu_k-\mu_l)^{{-4}/{3}},
\end{eqnarray}
where $C_0$ is an arbitrary constant, and $f_0$ is a value of $f$
corresponding to the seed $g_0$. In general, these solutions have two angular momentum component. In this article, we study the relationship between the solutions generated by the inverse scattering method applied to five-dimensions and those generated by the B\"acklund transformation in Ref.~\cite{MI} which generates the only solutions with a single angular momentum component. So, we should compare the solutions which have a single angular momentum component generated by them with each other. As discussed in Ref.~\cite{Tomizawa}, since the the two-solitonic solution (\ref{eq:nor}) with $n=2$ would not be regular on a certain part of an axis (as far as we choose a seed regular on it), it is suitable to normalize the metric so that
$(g_0)_{33}$ is unchanged:
\begin{eqnarray}
g^{{\rm (phys)}}=\left(
\begin{array}{@{\,}c|ccc@{\,}}
\displaystyle \left(\prod_{k=1}^n
\frac{\mu_k}{\rho}\right)
g^{{\rm (unphys)}}_{AB}
& 0 \\ \hline 0  &
(g_0)_{33}
\label{eq:norm12}
\end{array}
\right),
\end{eqnarray}
where $A, B=1,2$.
Here, we consider the two-soliton solution.
 We choose the sign of plus in Eq. (\ref{eq:mup})
and take the constants $w_1=-w_2=-\sigma $.


\section{Relation between two-solitonic solutions }\label{sec:proof}

In this section, we show that for a general diagonal seed solution which takes the form of
\begin{eqnarray}
ds^2=g'_1dt^2+g'_2d\phi^2+g'_3d\psi^2+f'(d\rho^2+dz^2),\label{eq:seed}
\end{eqnarray}
where $g'_1$, $g'_2$ and $g'_3$ are functions of $\rho$ and $z$, and satisfy the constraint $g'_1g'_2g'_3=-\rho^2$, the two-solitonic solutions generated by the inverse scattering method under the special normalization (\ref{eq:norm12}) coincide with ones generated by the B\"acklund transformation explained in Sec.\ref{sec:back}. 
To do so, as a diagonal seed, instead of (\ref{eq:seed}) it is sufficient to consider the following metric form,
\begin{eqnarray}
ds^2=-dt^2+g_2d\phi^2+g_3d\psi^2+f(d\rho^2+dz^2),\label{g23}
\end{eqnarray}
where $g_2$ and $g_3$ are functions of $\rho$ and $z$, and satisfy the constraint $g_2g_3=\rho^2$ (In fact, starting with this form of the seed metric symplfies the proof).

The reason for this is explained as follows. Let us consider the conformal transformation of the two dimensional metric $g_{AB}\ (A,B=t,\phi)$ and the rescale of the $\psi\psi$-component with the determinant ${\rm det} g$ invariant~;
\begin{eqnarray}
g_{0}={\rm diag} (-1,g_2,g_3)\to g'_0={\rm diag} (-\Omega,\Omega g_2,\Omega^{-2}g_3),\label{eq:conf}
\end{eqnarray} 
where $\ln\Omega$ must be a harmonic function on the three-dimensional Euclid space in order to assure that the transformed metric is the solution of Eq.(\ref{eq:soliton}). (Since the metric function $f$ or $f'$ is determined by only the three dimensional metric $g_0$ or $g'_0$, we need not consider this for the present purpose) Under this transformation, the physical metric (\ref{eq:norm12}) is transformed as
\begin{eqnarray}
g=\left(
\begin{array}{@{\,}c|ccc@{\,}}
\displaystyle
g_{AB}
& 0 \\ \hline 0  &
g_3
\end{array}
\right)
\to 
g^{\prime}=\left(
\begin{array}{@{\,}c|ccc@{\,}}
\displaystyle
\Omega g_{AB}
& 0 \\ \hline 0  &
\Omega^{-2}g_3
\end{array}
\right).\label{eq:tr}
\end{eqnarray}
 From this, we see that the transformation (\ref{eq:conf}) of a seed commutes 
with the operation of putting two-solitons on the background. 
Therefore we can obtain the two-solitonic solution generated from a diagonal seed
$g'_0={\rm diag}(g'_1,g'_2,g'_3)$ such that the $tt$-component is not $-1$
by the transformation (\ref{eq:tr}) with $\Omega=g'_1$
for the two-solitonic solution generated from the seed (\ref{g23}).

For the B\"acklund transformation, the same fact also holds, i.e. by the transformation (\ref{eq:conf}) of the seed, the solution generated is transformed as Eq.(\ref{eq:tr}).  
Note that the seed functions for the metric (\ref{g23})
can be written in terms of $g_2$ as
\begin{eqnarray}
S^{(0)}=T^{(0)}=-\frac{1}{2}\ln\biggl(\frac{g_2}{\rho^2}\biggr).\label{eq:st}
\end{eqnarray}
The two-solitonic solution for the general seed metric with seed functions
$S^{(0)}=-1/2\ln(g_2/\rho^2)$ and $T'_{(0)} \ne -1/2\ln(g_2/\rho^2)$
can be obtained from the solitonic solution of the seed (\ref{g23})
with the subsequent transformation (\ref{eq:conf}) 
with $\Omega=e^{S^{(0)}-T'_{(0)}}$.
As a result, we can conclude that it is sufficient to
assume the form of the diagonal seed as Eq.(\ref{g23}), i.e., the seed such that $(g_0)_{tt}=-1$ as far as we consider a diagonal seed solution.

To begin with, we show that for an arbitrary diagonal seed (\ref{g23}), the solutions of Eqs.(\ref{eq:ab}) are given by
\begin{eqnarray}
& & a=-\alpha(\kappa_1+1)\sigma^{3/2}\frac{g_2^{1/2}(x+1)(1-y)}{\rho\psi_2[\rho,z,\mu_2]},\quad b=-\beta\frac{\rho\psi_2[\rho,z,\mu_1]}{(\kappa_2-1)\sigma^{3/2}(x-1)(1-y)g_2^{1/2}},\label{eq:a1}
\end{eqnarray}
where $\psi_2[\rho,z,\lambda]$ is the $\phi\phi$-components of the solution $\psi$ of Eqs.(\ref{eq:Laxpair}) (we may assume the generating matrix $\psi_0[\rho,z,\lambda]$ to be diagonal $\psi_0[\rho,z,\lambda]={\rm diag}(\psi_1[\rho,z,\lambda],\psi_2[\rho,z,\lambda],\psi_3[\rho,z,\lambda])$ for a diagonal seed). $\alpha,\beta$, $\kappa_1$ and $\kappa_2$ are arbitrary constants.

 From Eq.(\ref{eq:st}), the right hand side in the first equation of (\ref{eq:ab}) is reduced to
\begin{eqnarray}
\frac{1}{x-y}[(xy-1)S^{(0)}_{,x}+(1-y^2)S^{(0)}_{,y}]=-\frac{1}{2(x-y)}\biggl[(xy-1)\biggl(\ln\frac{g_2}{\rho^2}\biggr)_{,x}+(1-y^2)\biggl(\ln\frac{g_2}{\rho^2}\biggr)_{,y}\biggr].\label{eq:right}
\end{eqnarray}
On the other hand, using the first equation of (\ref{eq:a1}),  the left hand side in Eq. (\ref{eq:ab}) becomes
\begin{eqnarray}
(\ln a)_{,x}&=&\frac{1}{2}(\ln g_2)_{,x}+\frac{1}{x+1}-(\ln\rho)_{,x}-(\ln\psi_2[\rho,z,\mu_2])_{,x}\nonumber\\
            &=&\frac{1}{2}(\ln g_2)_{,x}+\frac{1}{x+1}-(\ln\rho)_{,x}-\frac{\sigma^2}{\rho}x(1-y^2)(\ln\psi_2[\rho,z,\mu_2])_{,\rho}\nonumber\\
            & &-\sigma y(\ln\psi_2[\rho,z,\mu_2])_{,z}.\label{eq:lna}
\end{eqnarray}
Let us note that the term containing $(\ln\psi[\rho,z,\mu_2])_{,\rho}$ in the above equation is computed as
\begin{eqnarray}
(\ln\psi_2[\rho,z,\mu_2])_{,\rho}&=&(\ln\psi_2[\rho,z,\lambda])_{,\rho}|_{\lambda=\mu_2}+(\ln\psi_2[\rho,z,\lambda])_{,\lambda}|_{\lambda=\mu_2}\cdot\mu_{2,\rho}\nonumber\\
                                 &=&-\frac{2\mu_2\rho}{\rho^2+\mu_2^2}(\ln\psi_2[\rho,z,\lambda])_{,\lambda}|_{\lambda=\mu_2}+\frac{\rho^2(\ln g_2)_{,\rho}+\rho\mu_2(\ln g_2)_{,z}}{\rho^2+\mu_2^2}\nonumber\\
                                 &+&(\ln\psi_2[\rho,z,\lambda])_{,\lambda}|_{\lambda=\mu_2}\cdot \frac{2\mu_2\rho}{\rho^2+\mu_2^2}\nonumber\\
                                 &=&\frac{\rho^2(\ln g_2)_{,\rho}+\rho\mu_2(\ln g_2)_{,z}}{\rho^2+\mu_2^2}\label{eq:psirho}
\end{eqnarray}                                 
where we used Eq.(\ref{eq:Laxpair}) and (\ref{eq:eqmuk}). Similarly, the term containing $(\ln\psi[\rho,z,\mu_2])_{,z}$ in Eq.(\ref{eq:lna}) can be computed as
\begin{eqnarray}
(\ln\psi_2[\rho,z,\mu_2])_{,z}&=&\frac{\rho^2(\ln g_2)_{,z}-\rho\mu_2(\ln g_2)_{,\rho}}{\rho^2+\mu_2^2}\label{eq:psiz}
\end{eqnarray}
Therefore, using Eqs.(\ref{eq:psirho}) and (\ref{eq:psiz}), Eq.(\ref{eq:lna}) becomes
\begin{eqnarray}
(\ln  a)_{,x}=-\frac{1}{2(x-y)}\biggl[(xy-1)\biggl(\ln\frac{g_2}{\rho^2}\biggr)_{,x}+(1-y^2)\biggl(\ln\frac{g_2}{\rho^2}\biggr)_{,y}\biggr].
\end{eqnarray}
This coincides with the right hand side of Eq.(\ref{eq:right}), which implies that $a$ in Eq.(\ref{eq:a1}) is a solution of the first equation in Eq.(\ref{eq:ab}). Similarly, we can show $a$ satisfies the second equation of (\ref{eq:ab}) and that $b$ also satisfies the third and fourth equations of (\ref{eq:ab}). As a result we see that the solutions (\ref{eq:ab}) are given by Eqs.(\ref{eq:a1}) for a diagonal seed whose $tt$-component is $-1$.

Substituting Eq.(\ref{eq:a1}) into Eqs.(\ref{eq:sol}) and (\ref{eq:A-C}), we obtain the following general solution generated from a static seed solution,
\begin{eqnarray}
g_{tt}=-\frac{\tilde A}{\tilde B}, \quad
g_{t\phi}=2\sigma^{\frac{1}{2}}g_2\frac{\tilde C}{\tilde B}
+C_1\frac{\tilde A}{\tilde B},\quad
g_{\phi\phi}=\frac{g_{t\phi}^2-\rho^2}{g_{tt}}.\label{eq:MI}
\end{eqnarray}
Here we have introduced new functions
$\tilde A,\ \tilde B$ and $\tilde C$ defined as

\begin{eqnarray}
\tilde A&=&-\beta^2\psi_2[\mu_1]^2\psi_2[\mu_2]^2(1+y)^2
+\sigma\alpha^2\beta^2(\kappa_1+1)^2g_2\psi_2[\mu_1]^2(x+1)^2
\nonumber\\
& &+\sigma(\kappa_2-1)^2g_2\psi_2[\mu_2]^2(x-1)^2-\sigma^2\alpha^2(\kappa_1+1)^2(\kappa_2-1)^2g_2^2(1-y)^2\nonumber\\
& &+2\sigma\alpha\beta(\kappa_1+1)(\kappa_2-1)g_2\psi_2[\mu_1]
\psi_2[\mu_2](x^2-y^2),
\end{eqnarray}
\begin{eqnarray}
\tilde B&=&\beta^2\psi_2[ \mu_1]^2\psi_2[\mu_2]^2(1-y^2)
+\sigma^2\alpha^2(\kappa_1+1)^2(\kappa_2-1)^2g_2^2(1-y^2)\nonumber\\
& &+\sigma\alpha^2\beta^2(\kappa_1+1)^2\psi_2[\mu_1]^2g_2(x^2-1)
 +\sigma(\kappa_2-1)^2\psi_2[\mu_2]^2g_2(x^2-1)\nonumber\\
& &+2\sigma\alpha\beta(\kappa_1+1)(\kappa_2-1)g_2
\psi_2[\mu_1]\psi_2[\mu_2](x^2-y^2),
\end{eqnarray}
\begin{eqnarray}
\tilde C&=&-\beta(\kappa_2-1)
\psi_2[\mu_1]\psi_2[ \mu_2]^2(x+y)
-\alpha\beta^2(\kappa_1+1)\psi_2[\mu_1]^2
\psi_2[\mu_2](x-y)
\nonumber
\\
& &+\sigma\alpha^2\beta(\kappa_2-1)(\kappa_1+1)^2g_2
\psi_2[\mu_1](x+y)+\sigma\alpha(\kappa_1+1)
(\kappa_2-1)^2g_2\psi_2[\mu_2](x-y).
\end{eqnarray}

Next, let us consider the solutions generated from the inverse scattering method.
Under the special normalization (\ref{eq:norm12}),
the two-soliton solution can be written 
in the following form:

\begin{eqnarray}
& &g^{{\rm (phys)}}_{tt}=
-\frac{G_{tt}}{\mu_1\mu_2\Sigma},\quad
g_{t\phi}^{{\rm (phys)}}
=-g_2\frac{(\rho^2+\mu_1\mu_2)
G_{t\phi}}{\mu_1\mu_2 \Sigma},\quad
g^{{\rm (phys)}}_{\phi\phi}=
-g_2\frac{G_{\phi\phi}}{\mu_1\mu_2\Sigma},
\label{eq:gphys}
\\
& &g^{{\rm (phys)}}_{\psi\psi}=g_3,\quad
g_{\phi\psi}^{{\rm (phys)}}=g_{t\psi}^{{\rm (phys)}}=0,
\end{eqnarray}
where the functions 
$G_{tt},\ G_{t\phi},\ G_{\phi\phi}$
and $\Sigma$ are defined as
\begin{eqnarray}
G_{tt}&=&-m_{01}^{(1)2}m_{01}^{(2)2}
\psi_2[\mu_1]^2\psi_2[\mu_2]^2(\mu_1-\mu_2)^2
\rho^4+m_{01}^{(1)2}m_{02}^{(2)2}g_2\mu_2^2
(\rho^2+\mu_1\mu_2)^2\psi_2[\mu_1]^2
\nonumber\\
& &+m_{01}^{(2)2}m_{02}^{(1)2}g_2 \mu_1^2
(\rho^2+\mu_1\mu_2)^2\psi_2[\mu_2]^2
-m_{02}^{(1)2}m_{02}^{(2)2}g_2^2
\mu_1^2\mu_2^2(\mu_1-\mu_2)^2\\
& &-2m_{01}^{(1)}m_{01}^{(2)}m_{02}^{(1)}m_{02}^{(2)}
g_2\psi_2[\mu_1]\psi_2[\mu_2](\rho^2+\mu_1^2)
(\rho^2+\mu_2^2)\mu_1\mu_2,
\nonumber
\end{eqnarray}
\begin{eqnarray}
G_{\phi\phi}&=&m_{01}^{(1)2}m_{01}^{(2)2}
\mu_1^2\mu_2^2(\mu_1-\mu_2)^2
\psi_2[\mu_1]^2\psi_2[\mu_2]^2
+m_{02}^{(1)2}m_{02}^{(2)2}g_2^2
(\mu_1-\mu_2)^2\rho^4\nonumber\\
& &-m_{01}^{(1)2}m_{02}^{(2)2}g_2\mu_1^2
\psi_2[\mu_1]^2(\rho^2+\mu_1\mu_2)^2
-m_{01}^{(2)2}m_{02}^{(1)2}g_2\mu_2^2
(g_2-\mu_2)^2(\rho^2+\mu_1\mu_2)^2  \\
& &+2m_{01}^{(1)}m_{01}^{(2)}m_{02}^{(1)}m_{02}^{(2)}
g_2\mu_1\mu_2\psi_2[\mu_2]\psi_2[\mu_1]
(\rho^2+\mu_1^2)(\rho^2+\mu_2^2),
\nonumber
\end{eqnarray}
\begin{eqnarray}
G_{t\phi}&=&m_{01}^{(1)}m_{01}^{(2)2}
m_{02}^{(1)}\mu_2(\mu_1-\mu_2)
\psi_2[\mu_2]^2\psi_2[\mu_1](\rho^2+\mu_1^2)\nonumber\\
& &+m_{01}^{(1)}m_{02}^{(1)}m_{02}^{(2)2}
g_2\mu_2(\mu_2-\mu_1)
\psi_2[\mu_1](\rho^2+\mu_1^2)\\
& &+m_{01}^{(1)2}m_{01}^{(2)}m_{02}^{(2)}
\mu_1(\mu_2-\mu_1)\psi_2[\mu_1]^2
\psi_2[\mu_2](\rho^2+\mu_2^2)\nonumber\\
& &+m_{01}^{(2)}m_{02}^{(1)2}m_{02}^{(2)}
\mu_1g_2\psi_2[\mu_2](\rho^2+\mu_2^2)(\mu_1-\mu_2),\nonumber
\end{eqnarray}
\begin{eqnarray}
\Sigma&=&m_{01}^{(1)2}m_{01}^{(2)2}
\psi_2[\mu_1]^2\psi_2[\mu_2]^2(\mu_1-\mu_2)^2\rho^2
+m_{02}^{(1)2}m_{02}^{(2)2}g_2^2(\mu_1-\mu_2)^2\rho^2\nonumber\\
& &+m_{01}^{(1)2}m_{02}^{(2)2}g_2\psi_2[\mu_1]^2
(\rho^2+\mu_1\mu_2)^2
+m_{02}^{(1)2}m_{01}^{(2)2}g_2
\psi_2[\mu_2]^2(\rho^2+\mu_1\mu_2)^2\nonumber\\
& &-2m_{01}^{(1)}m_{01}^{(2)}m_{02}^{(1)}m_{02}^{(2)}
g_2\psi_2[\mu_1]\psi_2[\mu_2](\rho^2+\mu_1^2)(\rho^2+\mu_2^2),
\end{eqnarray}
where the two functions $g_2$ and $g_3$ are given by Eq. (\ref{g23}).

In order for the metric to approach the Minkowski spacetime
asymptotically,
let us consider the coordinate transformation
of the physical metric such that
\begin{eqnarray}
t\rightarrow t'=t-C_1\phi, \qquad \phi
\rightarrow \phi'=\phi,
\end{eqnarray}
where $C_1$ is a constant chosen to ensure the
asymptotic flatness.
We should note that the transformed metric also satisfies the
supplementary condition ${\rm det} g=-\rho^2$.
Under this transformation, the physical metric components become
\begin{eqnarray}
& &g_{tt}^{\rm (phys)}\rightarrow
g_{t't'}^{\rm (phys)}=g_{tt}^{\rm (phys)},
\nonumber \\
& &g_{t\phi}^{\rm (phys)}\rightarrow
g_{t'\phi'}^{\rm (phys)}=g_{t\phi}^{\rm (phys)}+C_1
g_{tt}^{\rm (phys)},\\
& &g_{\phi\phi}^{\rm (phys)}\rightarrow
g_{\phi'\phi'}^{\rm (phys)}=g_{\phi\phi}^{\rm (phys)}
+2C_1 g_{t\phi}^{\rm (phys)}+C_1^2g_{tt}^{\rm (phys)}.
\nonumber
\end{eqnarray}
If we choose the parameters such that
\begin{eqnarray}
m_{01}^{(1)}m_{01}^{(2)}&=&\beta,\\
m_{01}^{(2)}m_{02}^{(1)}&=&\sigma^{\frac{1}{2}}(\kappa_2-1),\\
m_{01}^{(1)}m_{02}^{(2)}&=&-\sigma^{\frac{1}{2}}\alpha\beta(\kappa_1+1),\\
m_{02}^{(1)}m_{02}^{(2)}&=&-\sigma\alpha(\kappa_1+1)(\kappa_2-1), \\
C_1 &=&\frac{2\sigma ^{1/2}\alpha }{1+\alpha \beta },
\end{eqnarray}
and use the prolate spherical coordinate $(x,y)$, we can confirm that the transformed metric coincides with the metric (\ref{eq:MI}) generated by the technique used in Ref.~\cite{MI} from a diagonal seed. In order to show the coincidence of the metrics, it is sufficient to check only two components $g_{tt}$ and $g_{t\phi}$ due to the supplementary condition ${\rm det} g=-\rho^2$ and the fact that the metric components $g_{\rho\rho}$ and $g_{zz}$ (or, $g_{xx}$ and $g_{yy}$) are determined by the three-dimensional metric $g_{ij}$.


\section{Summary and Discussion}

In this article, we studied the relation between the inverse scattering method and the B\"acklund transformation applied to five dimensions. We showed that the two-solitonic solution generated from an arbitrary diagonal seed by the B\"acklund transformation coincides with one generated from the same diagonal seed by the inverse scattering method under the special normalization (\ref{eq:norm12}).
This implies that the five-dimensional solutions generated by the inverse scattering method contain the ones generated by the B\"acklund transformation used in Ref.~\cite{MI} as concerned with the two-solitonic solutions generated from a diagonal seed. As clarified in the previous works~\cite{MI,Tomizawa,MI2,Tomizawa2}, if we choose the five-dimensional Minkowski or the Euclidean $C$-metric as a diagonal seed, we can obtain the black ring solution with a rotating two-sphere~\cite{MI} or the black ring solution found by Emparan and Reall~\cite{Emparan:2001wn} as the two-solitonic solution, respectively. Therefore, 
we see that the previous works~\cite{Tomizawa,Tomizawa2} correspond to the special cases of the present result.

However, while the B\"acklund transformation used in Ref.~\cite{MI} can 
generate solutions with a single angular momentum component at the most, the inverse scattering method can generate five-dimensional solutions with two angular momentum components, as discussed in Ref.~\cite{Tomizawa,Tomizawa2}. In fact, the Myers-Perry black hole solution with two angular momentum components was generated by the inverse scattering method~\cite{Pomeransky:2005sj}. It is expected that a new black ring solution with two angular momentum components may be generated by this method.

Though in this article, we focus on the two-solitonic solution generated from a diagonal seed by these techniques, we also expect solutions generated from non-diagonal seeds by both generation techniques or multi-solitonic solution (more than two) generated by them to be same if we choose the special normalization (\ref{eq:norm12}) in the inverse scattering method. It would be also interesting to deal with the generation of solutions in the five-dimensional Einstein-Maxwell equations with the same symmetries~\cite{E}.

\section*{Acknowledgments}
We thank Ken-ichi~Nakao and Hideki~Ishihara for continuous encouragement.
HI is supported in part by Grant-in-Aid for Young Scientists (B)
(No. 17740152) from Japanese Ministry of Education, Science,
Sports, and Culture.


\begin{thebibliography}{22}

\bibitem{BHinCollider}
T.~Banks and W.~Fischler,
arXiv:hep-th/9906038;
S.~B.~Giddings and S.~D.~Thomas,
Phys.\ Rev.\ D {\bf 65}, 056010 (2002);
S.~Dimopoulos and G.~Landsberg,
Phys.\ Rev.\ Lett.\  {\bf 87}, 161602 (2001).

\bibitem{Topology} 
M.~I.~Cai and  G.~J.~Galloway, Class. Quant. Grav. {\bf 18}, 2707 (2001);
C. Helfgott, Y.Oz, and Y. Yanay, JHEP, {\bf 02}, 025 (2006);
G. J. Galloway and R.Schoen, arXive:gr-qc/0509107.



\bibitem{Myers:1986un}
R.~C.~Myers and M.~J.~Perry,
Annals Phys.\  {\bf 172}, 304 (1986).



\bibitem{Emparan:2001wn}
R.~Emparan and H.~S.~Reall, Phys.\ Rev.\ Lett.\  {\bf 88}, 101101 (2002);
R.~Emparan and H.~S.~Reall, arXive:hep-th/0608012. 


\bibitem{MI}
T.~Mishima and H.~Iguchi, Phys. Rev. D{\bf 73}, 044030, (2006);
H.~Iguchi and T.~Mishima, Phys. Rev. D {\bf 74}, 024029 (2006).



\bibitem{Castejon-Amenedo:1990b}
J.~Castejon-Amenedo and V.~S.~Manko, Phys.\ Rev.\ D {\bf 41}, 2018 (1990).

\bibitem{MI2}
H.Iguchi and T.Mishima, Phys. Rev. D {\bf 73}, 121501(R) (2006).

\bibitem{Belinskii}
V.~A.~Belinskii and V.~E.~Zakharov, Sov. Phys. JETP {\bf 50}, 1 (1979);
V.~A.~Belinskii and V.~E.~Zakharov, Sov. Phys. JETP {\bf 48}, 985 (1978).








\bibitem{textism}
V.~A.~Belinski and E.~Verdaguer, {\it Gravitational Solitons} (CambridgeUniversity Press, Cambridge, England, 2001).



\bibitem{exact}
H.~Stephani, D.~Kramer, M.~MacCallum, C.~Hoenselaers and E.~Herlt, {\it Exact solutions of Einstein's Field Equations, 2nd ed.} (Cambridge University Press, Cambridge, 2003).



\bibitem{Koikawa:2005ia}
T.~Koikawa, Prog.\ Theor.\ Phys.\  {\bf 114}, 793 (2005).






\bibitem{Tomizawa}
S. Tomizawa, Y. Morisawa, Y Yasui, Phys. Rev. D{\bf 73}, 064009 (2006).


\bibitem{G}
M. G{\" u}rses, {\it Inverse Scattering, Differential
Geometry, Einstein Maxwell Solitons and One Soliton Backlund
Transformations} in {\it Solutions of Einstein's Equations:
Techniques and Results} Edited by C. Hoenselaers and W. Dietz,
(Springer-Verlag , Berlin, 1984).


\bibitem{G1}
M. G{\" u}rses, Phys.Rev.Lett. {\bf 51}, 1810 (1983);
M. G{\" u}rses, Phys.Rev.D {\bf 30}, 486(1984).
\bibitem{Cos}
C. M. Cosgrove, J.Math.Phys. {\bf 21}, 2417 (1980).

\bibitem{exact}
H.~Stephani, D.~Kramer, M.~MacCallum, C.~Hoenselaers and E.~Herlt,
{\it Exact solutions of Einstein's Field Equations, 2nd ed.}
(Cambridge University Press, Cambridge, 2003).

\bibitem{Pomeransky:2005sj}
A.~A.~Pomeransky, Phys. Rev. D {\bf 73}, 044004 (2006). 

\bibitem{Tomizawa2}
S. Tomizawa and M. Nozawa, Phys. Rev. D {\bf 73}, 124034 (2006). 


\bibitem{refERNST} 
F.~J.~Ernst, Phys. Rev. {\bf 167}, 1175 (1968).



\bibitem{Neugebauer:1980}
G.~Neugebauer, J.\ Phys.\ A {\bf 13}, L19 (1980).



\bibitem{Hoenselaers:1979mk}
C.~Hoenselaers, W.~Kinnersley and B.~C.~Xanthopoulos, J.\ Math.\ Phys.\  {\bf 20}, 2530 (1979).


\bibitem{ref7} 
J.~Castejon-Amenedo and V. S.~Manko, Phys. Rev. D {\bf 41}, 2018 (1990).


\bibitem{weyl}
R. Emparan, H. S. Reall, Phys. Rev. D {\bf 65}, 084025 (2002). 





\bibitem{E}
A. Eri\c{s}, M. G{\" u}rses and A. Karasu, J.Math.Phys.
{\bf 25}, 1489(1984).
 

\end{thebibliography}
\end{document}